\documentstyle[preprint,aps,epsf,floats]{revtex}
\begin{document}
\tighten
\def\si{{}^1\kern-.14em S_0}
\def\siii{{}^3\kern-.14em S_1}
\def\diii{{}^3\kern-.14em D_1}
\def\pone{{}^3\kern-.14em P_1}
\def\pzero{{}^3\kern-.14em P_0}
\def\ptwo{{}^3\kern-.14em P_2}
\newcommand{\gsim}{\raisebox{-0.7ex}{$\stackrel{\textstyle >}{\sim}$ }}
\newcommand{\lsim}{\raisebox{-0.7ex}{$\stackrel{\textstyle <}{\sim}$ }}
\def\pislash{ {\pi\hskip-0.6em /} }
\def\pislashsmall{ {\pi\hskip-0.375em /} }
\def\pslash{p\hskip-0.45em /}
\def\nopi{ {\rm EFT}(\pislash) }
\def\Ltwo{ {^\pislashsmall \hskip -0.2em L_2^{(M1)} }}
\def\Lone{ {^\pislashsmall \hskip -0.2em L_1^{(M1)} }}
\def\LEone{ {^\pislashsmall \hskip -0.2em L_1^{(E1)} }}
\def\LEthree{ {^\pislashsmall \hskip -0.2em L_3^{(E1)} }}
\def\CQuad{ {^\pislashsmall \hskip -0.2em C_{\cal Q} }}
\def\Czeromone{ {^\pislashsmall \hskip -0.2em C_{0,-1}^{(\siii)} }}
\def\Czerozero{ {^\pislashsmall \hskip -0.2em C_{0,0}^{(\siii)} }}
\def\Czeroone{ {^\pislashsmall \hskip -0.2em C_{0,1}^{(\siii)} }}
\def\Ctwomtwo{ {^\pislashsmall \hskip -0.2em C_{2,-2}^{(\siii)} }}
\def\Ctwomone{ {^\pislashsmall \hskip -0.2em C_{2,-1}^{(\siii)} }}
\def\Ctwomone{ {^\pislashsmall \hskip -0.2em C_{2,-1}^{(\siii)} }}
\def\CSDzero{ {^\pislashsmall \hskip -0.2em C_0^{(sd)} }}
\def\CSDtwotwotwo{ {^\pislashsmall \hskip -0.2em C_{2,-2}^{(sd)} }}
\def\CSDzeromone{ {^\pislashsmall \hskip -0.2em C_{0,-1}^{(sd)} }}
\def\CSDzerozero{ {^\pislashsmall \hskip -0.2em C_{0,0}^{(sd)} }}
\def\CSDtwoone{ {^\pislashsmall \hskip -0.2em \tilde C_2^{(sd)} }}
\def\CSDtwotwo{ {^\pislashsmall \hskip -0.2em C_2^{(sd)} }}
\def\CSDzerotwo{ {^\pislashsmall \hskip -0.2em C_{0,0}^{(sd)} }}
\def\LX{ {^\pislashsmall \hskip -0.2em L_X }}
\def\Czero{ {^\pislashsmall \hskip -0.2em C_0^{(\siii)} }}
\def\Ctwo{ {^\pislashsmall \hskip -0.2em C_2^{(\siii)} }}
\def\Cfourone{ {^\pislashsmall \hskip -0.2em \tilde C_4^{(\siii)} }}
\def\Cfouronemtwo{ {^\pislashsmall \hskip -0.2em \tilde C_{4,-2}^{(\siii)} }}
\def\Cfourtwo{ {^\pislashsmall \hskip -0.2em C_4^{(\siii)} }}
\def\Csixtwo{ {^\pislashsmall \hskip -0.2em  C_6^{(\siii)} }}
\def\Czerominussing{ {^\pislashsmall \hskip -0.2em C_{0,-1}^{(\si)} }}
\def\Ctwomtwosing{ {^\pislashsmall \hskip -0.2em C_{2,-2}^{(\si)} }}
\def\Csixtwomfour{ {^\pislashsmall \hskip -0.2em  C_{6,-4}^{(\siii)} }}
\def\CPzero{ {^\pislashsmall \hskip -0.2em C^{(\pzero)}_2  }}
\def\CPone{ {^\pislashsmall \hskip -0.2em C^{(\pone)}_2  }}
\def\CPtwo{ {^\pislashsmall \hskip -0.2em C^{(\ptwo)}_2  }}

\def\Lnp{ {^\pislashsmall \hskip -0.2em L_{np} }}

\def\Tr{{\rm Tr\,}}

% A useful Journal macro
\def\Journal#1#2#3#4{{#1} {\bf #2}, #3 (#4)}

% Some useful journal names
\def\NCA{\em Nuovo Cimento}
\def\NIM{\em Nucl. Instrum. Methods}
\def\NIMA{{\em Nucl. Instrum. Methods} A}
\def\NPB{{\em Nucl. Phys.} B}
\def\NPA{{\em Nucl. Phys.} A}
\def\NP{{\em Nucl. Phys.} }
\def\PLB{{\em Phys. Lett.} B}
\def\PRL{\em Phys. Rev. Lett.}
\def\PRD{{\em Phys. Rev.} D}
\def\PRC{{\em Phys. Rev.} C}
\def\PRA{{\em Phys. Rev.} A}
\def\PR{{\em Phys. Rev.} }
\def\ZPC{{\em Z. Phys.} C}
\def\SJP{{\em Sov. Phys. JETP}}
\def\SJNP{{\em Sov. Phys. Nucl. Phys.}}

\def\FBS{{\em Few Body Systems Suppl.}}
\def\IJMP{{\em Int. J. Mod. Phys.} A}
\def\UJP{{\em Ukr. J. of Phys.}}
\def\CJP{{\em Can. J. Phys.}}
\def\SCI{{\em Science} }
\def\AST{{\em Astrophys. Jour.} }

\preprint{\vbox{
\hbox{ NT@UW-99-35}
}}
\bigskip
\bigskip

\title{$np\rightarrow d\gamma$ for Big-Bang Nucleosynthesis}
\author{Jiunn-Wei Chen$^a$, and Martin J. Savage$^{a,b}$}
\address{$^a$ Department of Physics, University of Washington, \\
Seattle, WA 98915. }
\address{$^b$ Jefferson Lab., 12000 Jefferson Avenue, Newport News, \\
Virginia 23606.}
\maketitle

\begin{abstract}
The cross section for  $np\rightarrow d\gamma$ is calculated 
at energies  relevant to big-bang nucleosynthesis using 
the recently developed effective field theory
that describes the two-nucleon sector.
The E1 amplitude is computed up to N$^3$LO and depends only upon
nucleon-nucleon phase shift data.
In contrast, the M1 contribution is computed up to NLO, and the 
four-nucleon-one-magnetic-photon counterterm that enters is determined 
by the cross section for cold neutron capture.
The uncertainty in the calculation for nucleon energies up 
to $E\sim 1~{\rm MeV}$ is estimated to be $\lsim 4\%$.
\end{abstract}

\vskip 2in

\leftline{July, 1999}
%%%%%%%%%%%%%%%%%%%%%%%%%%%%%%%%%
%
%        VERSION DATE
%
% \leftline{{\bf Draft version sometime 1999}}
%
%
%%%%%%%%%%%%%%%%%%%%%%%%%%%%%%%%%%
\vfill\eject

%%%%%%%%%%  Intro %%%%%%%%%%%%%%%%

The radiative capture process $np\rightarrow d\gamma$ is a key reaction
in the synthesis of nuclei in the early universe.
Recently, it has been emphasized by
Burles, Nollet, Truran and Turner (BNTT)\cite{BNTTa}
that the uncertainty in the cross section\cite{ENDF}
of $np\rightarrow d\gamma$ at energies relevant for 
big-bang nucleosynthesis (BBN) is difficult to determine
due to the lack of data at low energies
and the lack of information about theoretical estimates.
In determining theoretical uncertainties
in the abundances of the elements produced in 
BBN,
Smith, Kawano and Malaney (SKM)\cite{SKMa} assigned a $1\sigma$ 
error of $5\%$ to the cross section for $np\rightarrow d\gamma$, 
which was also used in the recent analysis of 
BNTT\cite{BNTTa}.
In \cite{BNTTa} it was found that this $5\%$ uncertainty 
contributes a significant fraction of the 
uncertainties in the abundances of elements
produced in BBN.

In this work we compute the cross section of $np\rightarrow d\gamma$
using the recently developed techniques of effective field theory in the 
two-nucleon sector\cite{KSW,KSW2}.
For the energy range appropriate for BBN 
(nucleon energies $E_N\lsim 1 {\rm MeV}$)
it is appropriate to use the effective field theory of only 
nucleons and photons, as presented in \cite{CRSa},
which we denote by $\nopi$.
The cross section of $np\rightarrow d\gamma$ for cold neutrons has been 
computed in the theory with pions~\cite{SSWst} and in 
$\nopi$~\cite{CRSa} up to next-to-leading order (NLO) in the 
effective field theory expansion parameter(s).
For cold neutrons the cross section is dominated by 
M1-capture from the $\si$ channel
via the nucleon isovector magnetic moment.
However, in addition to the contribution from the effective ranges
of both the $\si$ and $\siii$ channels at NLO,
there is a contribution from a 
four-nucleon-one-magnetic-photon interaction with a coefficient, 
$\Lone$, that is not 
constrained by nucleon-nucleon scattering phase shift data.
The observed cross section for cold neutrons determines $\Lone$.
At higher energies, $E_N\sim 1~{\rm MeV}$, the cross section for 
$np\rightarrow d\gamma$  is dominated by the
E1-capture of nucleons in a relative P-wave, 
$\pzero$, $\pone$, and $\ptwo$.
In the energy region relevant to BBN the contributions from both 
E1- and M1-capture are important.

It is important to emphasize that the results of our calculation
look very similar to those of effective range 
theory (ER) \cite{ERtheory,Noyes} 
(for a detailed discussion see \cite{ASbook}).
One of the interesting results from the recent developments in effective field
theory is that ER  is seen to reproduce the leading
orders of any particular amplitude for low energy processes.  
However, ER fails to reproduce the true
amplitude at and beyond the order at which there is a contribution from a
local, multi-nucleon-external-field interaction~\cite{CRSa}.
If multi-nucleon-external-field interactions do not enter until very high
orders in the $\nopi$  expansion, then ER will
reproduce the observed value to high precision, as is the case for the
polarizability of the deuteron\cite{CRSa,CGSSpol}.
However, if a multi-nucleon-external-field interaction occurs at low orders ER
can deviate substantially from the true result, as is the case for the 
the capture of cold neutrons, $np\rightarrow d\gamma$\cite{SSWst}, or the
deuteron quadrupole moment\cite{KSW2}.
For this process, the conventional understanding of this discrepancy is that
important contributions from meson-exchange currents have been 
omitted\cite{RiB}.
However, in effective field theory, this discrepancy results from 
the omission of  
four-nucleon-one-magnetic-photon operators 
that enter  at NLO and higher in the expansion.
For $np\rightarrow d\gamma$ at finite but low 
incident nucleon energy, the two dominant amplitudes, E1 and M1, behave
differently in the effective field theory expansion.
In $\nopi$, a four-nucleon-one-electric-dipole-photon local operator
occurs at N$^4$LO, which means that the E1 amplitude can be computed up to 
N$^3$LO with knowledge of only the nucleon-nucleon scattering phase shifts.
Therefore, this amplitude will look very similar to the expression obtained
in ER, when a $\gamma\rho$ expansion is performed.
In contrast, the M1 amplitude receives a contribution from a 
four-nucleon-one-magnetic-photon at NLO, and therefore the effective field
theory result will deviate from that obtained in ER in  a significant way.
In addition to the expressions we obtain for both the E1 and M1 amplitudes 
being analytic
and compact, they are perturbatively close to the true amplitudes for this
process,  giving a total cross section that deviates 
$\lsim 4\%$ over
the range of center-of-mass kinetic energies below $1~{\rm MeV}$.
In our calculation we neglect both isospin violation and 
relativistic effects, as in the energy region of interest both effects 
are significantly smaller than
the uncertainty introduced by not computing beyond the order to which 
we work
(relativistic effects are formally NNLO in $\nopi$\cite{CRSa}, 
but are suppressed by
additional factors of $m_\pi^2/M_N^2$ compared to other NNLO effects).

The strong interactions between two nucleons in the $\siii$-channel
are determined by the lagrange density, up to N$^3$LO,
\begin{eqnarray}
{\cal L}_{2}^{(\siii)} & = & 
- \Czero \left(N^T P_i N\right)^\dagger\left(N^T P_i N\right)
\ +\  {1\over 8} \Ctwo
\left[(N^T P_i N)^\dagger \left(N^T {\cal O}^{(2)}_i N\right)  +  h.c.\right]
\nonumber\\
 & - &  {1\over 16}\ \Cfourtwo\ \left( N^T {\cal O}^{(2)}_i  N\right)^\dagger
 \left( N^T{\cal O}^{(2)}_i N\right)
\ -\ 
{1\over 32}\ \Cfourone \left[ 
\left( N^T{\cal O}^{(4)}_i N\right)^\dagger
\left( N^T P_i N\right)\ +\ {\rm h.c.} \right]    
\nonumber\\
& + & 
{1\over 128} \ \Csixtwo 
\left[ 
\left( N^T{\cal O}^{(4)}_i N\right)^\dagger
\left( N^T {\cal O}^{(2)}_i N\right) \ +\ {\rm h.c.} \right]    
\label{eq:lagtwo}
\end{eqnarray}
where $P_i$ is the spin-isospin projector for the $\siii$ channel
\begin{eqnarray}
P_i \equiv {1\over \sqrt{8}} \sigma_2\sigma_i\ \tau_2
\ \ \ , 
\qquad \Tr P_i^\dagger P_j ={1\over 2} \delta_{ij}
\ \ \ .
\end{eqnarray}
The Galilean invariant
derivative operators ${\cal O}^{(2)}$ and ${\cal O}^{(4)}$ are defined by
\begin{eqnarray}
{\cal O}^{(2)}_i & = &  
P_i \overrightarrow {\bf D}^2 +\overleftarrow {\bf D}^2 P_i
    - 2 \overleftarrow {\bf D} P_i \overrightarrow {\bf D} 
\nonumber\\
{\cal O}^{(4)}_i & = &  P_i \overrightarrow {\bf D}^4 - 
4  \overleftarrow {\bf D} P_i \overrightarrow {\bf D}^3
+ 6  \overleftarrow {\bf D}^2 P_i \overrightarrow {\bf D}^2
- 4   \overleftarrow {\bf D}^3 P_i \overrightarrow {\bf D}
+ \overleftarrow {\bf D}^4 P_i
\ \ \ ,
\label{eq:ddef}
\end{eqnarray}
where the covariant derivative is defined to be 
${\bf D} = \nabla - ieQ{\bf A}$.
The superscript on the coefficient denotes the number of derivatives in the 
operator.
The expansion parameters of $\nopi$ are the external momentum involved in the 
particular process normalized to the mass of the pion, $Q\sim p/m_\pi$.
For momenta of order the pion mass or greater this expansion will fail 
to converge.
We have not shown the other operator involving six derivatives as it 
does not contribute to 
$np\rightarrow d\gamma$ at the order to which we are working.
In order that the deuteron pole is not shifted order-by-order in the 
$Q$-expansion,
the coefficients appearing in eq.~(\ref{eq:lagtwo})
have expansions in powers in $Q$ ,
e.g.
\begin{eqnarray}
    \Czero & = & \Czeromone + \Czerozero + \Czeroone\ +\ ...
\ \ \ .
\end{eqnarray}
The second subscript on each coefficient
denotes the powers of $Q$ in the coefficient itself.
Relating the S-matrix obtained from the lagrange density in
eq.~(\ref{eq:lagtwo}) to that described by the effective range expansion,
\begin{eqnarray}
  |{\bf k}|\cot\delta_0 & = & -\gamma_t \ +\
  {1\over 2}\rho_d (|{\bf k}|^2+\gamma_t^2)\ +\
  w_2\ (|{\bf k}|^2+\gamma_t^2)^2\ +\ ...
\ \ \ ,
\label{eq:kcot}
\end{eqnarray}
one can fix most of the coefficients 
(only one linear combination of $\Cfourtwo$ and $\Cfourone$ 
contributes to NN scattering)
appearing in 
eq.~(\ref{eq:lagtwo}) in terms of 
$\gamma_t^{-1} = 4.318946\ {\rm fm}$ 
($\gamma_t$ is the deuteron binding momentum),
$\rho_d = 1.764\pm 0.002\ {\rm fm}$ (the effective range parameter), 
and $w_2=0.389\ {\rm fm^3}$ (the shape parameter)\cite{Nij}.
The neglect of relativistic effects allows us to 
set $\gamma_t=\gamma = \sqrt{M_N B}$
where $B=2.224575~{\rm MeV}$ is the deuteron binding energy.
In addition, to the order we are working mixing between the $\siii$ and $\diii$
channels does not contribute, and so we will not discuss this sector.
However, there is a contribution from P-wave final state interactions in the 
E1-capture process that enter at N$^3$LO.
The P-wave interactions are described at leading order 
by the lagrange density
\begin{eqnarray}
{\cal L}_2^{P} & = & 
\left( \CPzero \delta^{xy}\delta^{wz}\ +\ 
\CPone \left[ \delta^{xw}\delta^{yz} - \delta^{xz}\delta^{yw}\right]
\ +\ 
\CPtwo \left[ 2 \delta^{xw}\delta^{yz} + 2 \delta^{xz}\delta^{yw}
- {4\over 3} \delta^{xy}\delta^{wz}\right]
\right)
\nonumber\\
& & 
\times\ 
{1\over 4}
\left( N^T
{\cal O}^{(1,P)}_{xy}  N\right)^\dagger
\left( N^T
{\cal O}^{(1,P)}_{wz}  N\right)
\ \ \ ,
\label{eq:lagp}
\end{eqnarray}
where the P-wave operators are
\begin{eqnarray}
{\cal O}^{(1,P)}_{ij} & = & 
 \overleftarrow {\bf D}_i P^{(P)}_j- P^{(P)}_j \overrightarrow {\bf D}_i
\ \ \ ,
\label{eq:Pops}
\end{eqnarray}
and $P^{(P)}_i$ is the spin-isospin projector for the isotriplet, spintriplet
channel
\begin{eqnarray}
P^{(P)}_i \equiv {1\over \sqrt{8}} \sigma_2\sigma_i\ \tau_2\tau_3
\ \ \ , 
\qquad \Tr P^{(P)\dagger}_i P^{(P)}_j ={1\over 2} \delta_{ij}
\ \ \ .
\end{eqnarray}
The measured P-wave phase shifts (as given by the Nijmegen phase shift 
analysis\cite{nijmegen})
fix the coefficients appearing in
eq.~(\ref{eq:lagp}) to be 
\begin{eqnarray}
\CPzero & = & +6.53\ {\rm fm}^4
\qquad , \qquad
\CPone \ =\  -5.91\ {\rm fm}^4
\qquad {\rm and } \qquad
\CPtwo \ =\  +0.57\ {\rm fm}^4
\ \ \ .
\label{eq:pcoeff}
\end{eqnarray}

Finally, there are interactions with the electromagnetic field that are not
simply related by gauge invariance to the strong interaction dynamics.
The lagrange density describing the leading interactions that contribute to 
$np\rightarrow d\gamma$ is
\begin{eqnarray}
{\cal L}_{2,B}
& = &  \left[
e\  \Lone \ (N^T\  P_i \ N)^\dagger (N^T\  \overline{P}_3 \ N)\  {\bf B}_i
+ {\rm h.c.}
\right.
\nonumber\\
& & \left. 
\ +\ 
{1\over 2}\ e\  \LEone (N^T \ {\cal O}^{(1,P)}_{ia} \ N)^\dagger 
(N^T \ P_a \ N) \ {\bf E}_i
+ {\rm h.c.}
\right.
\nonumber\\
& & \left. 
\ -\ {1\over 8} \ e\ \LEthree
(N^T \ {\cal O}^{(1,P)}_{ia} \ N)^\dagger 
(N^T\  {\cal O}^{(2)}_a  \ N) \ {\bf E}_i
+ {\rm h.c.}
\right]
\ \ \ \ ,
\label{eq:Ldef}
\end{eqnarray}
where  ${\bf E}$ is the electric field,
${\bf B} ={\bf \nabla} \times {\bf A}$ is the magnetic field, and 
where 
\begin{eqnarray}
\overline{P}_3 & = & {1\over\sqrt{8}} \sigma_2\ \tau_2\tau_3
\ \ \ ,
\end{eqnarray}
is the projector for the $\si$ channel.
The renormalization group (RG) evolution of 
$\Lone$ has been discussed in \cite{SSWst}, and the 
evolution of the electric coefficients are
determined by
\begin{eqnarray}
    \mu {d\over d\mu}
  \left[  {\LEone\ -\ M_N \Cfouronemtwo \over \Czeromone}\right]
 & = & 0
\qquad ,\qquad 
   \mu {d\over d\mu}
  \left[  {\LEthree\ -\ M_N \Csixtwomfour \over \Czeromone}\right]
 \ =\  0
\ \ \ .
\label{eq:LoneRG}
\end{eqnarray}
These RG equations tell us (by considering the size of the quantities
in square brackets at $\mu=m_\pi$, the matching scale)
that the combination
$\LEone - M_N \Cfouronemtwo$ is of order $Q^{-1}$ or higher,
as opposed to the naive counting of $Q^{-2}$ and that
$\LEthree - M_N \Csixtwomfour$
is of order $Q^{-1}$ or higher, as opposed to $Q^{-4}$.

%%%%%%%%%%%%  Amplitude  %%%%%%%%%%%%%%%%

The amplitude for $np\rightarrow d\gamma$ can be written as
\begin{eqnarray}
  T & = &
  e\  X_{E1}\  U^{T}_{\rm n}\ \tau_2\tau_3\ {\bf \sigma }_2\ 
  {\bf\sigma}\cdot\epsilon _{(d)}^*
\ U_{\rm p}
\ \ {\bf P}\cdot\epsilon_{(\gamma)}^*
\ +\  i e \ X_{M1}\ \varepsilon ^{abc}\epsilon _{(d)}^{\ast a}\
{\bf k}^{b}\
\epsilon _{(\gamma )}^{\ast c}\ 
U^{T}_{\rm n} \tau _{2}\tau _{3}\ {\bf \sigma }_{2}\ U_{\rm p} 
\ \ \ ,
\label{eq:npamp}
\end{eqnarray}
where we have not shown amplitudes that contribute much less than
$1\%$ to the total capture cross section in the energy range of interest,
leaving only the isovector E1 and the isovector M1 amplitudes.
$U_{\rm n}$ is the neutron two-component spinor and $U_{\rm p}$ 
is the proton two-component spinor.
$|{\bf P}|$ is the magnitude of the 
momentum of each nucleon in the center of mass frame,
while ${\bf k}$ is the photon momentum.
The  photon polarization vector is 
$\epsilon _{(\gamma )}$, and $\epsilon _{(d)}$ is the deuteron polarization
vector. For convenience, we define dimensionless variables $\tilde{X}$, by 
\begin{eqnarray}
\frac{|{\bf P}| M_{N}}{\gamma ^{2}}X_{E1}\  &=&i\frac{2}{M_{N}}
\sqrt{\frac{\pi }
  {\gamma ^{3}}}\
\tilde{X}_{E1}\quad ,\quad
X_{M1}\ =i\frac{2}{M_{N}}
\sqrt{\frac{\pi }{\gamma ^{3}}}\ \tilde{X}_{M1}
\quad ,
\label{eq:tildedef}
\end{eqnarray}
In terms of these amplitudes, the total cross section for 
$np\rightarrow d\gamma$ is 
\begin{eqnarray}
 \sigma & = & {4\pi\alpha \left(\gamma^2+|{\bf P}|^2\right)^3
\over \gamma^3 M_N^4 |{\bf P}|}
\left[\  |\tilde{X}_{M1}|^{2}
  \ +\ |\tilde{X}_{E1}|^{2}\ 
\right]
\ \ \ ,
\label{unpol}
\end{eqnarray}
where we have used nonrelativistic kinematics, as is appropriate for 
the energy region of interest.

%%%%%%%%%  E1  %%%%%%%%%%%%%%%%%%

Explicit calculation of $|\tilde{X}_{E1}|^2$ up to N$^3$LO (i.e. $Q^1$) 
gives
\begin{eqnarray}
 |\tilde{X}_{E1}|^2 & = & 
{|{\bf P}|^2 M_N^2 \gamma^4 \over \left(\gamma^2+|{\bf P}|^2\right)^4}
\left[ 1 + \gamma \rho_d + (\gamma \rho_d)^2 + (\gamma \rho_d)^3 
\right.
\nonumber\\
& & \left.\qquad\qquad
\ +\ {M_N\gamma\over 6\pi}
\left({\gamma^2\over 3}+|{\bf P}|^2\right)
\left( \CPzero + 2 \  \CPone + {20\over 3}\  \CPtwo \right)
\right]
\ \ \ .
\label{eq:Eone}
\end{eqnarray}
The momentum expansion of the theory is made explicit in eq.~(\ref{eq:Eone}),
and it is clear that we have captured all terms up to and including 
$\left(\gamma/m_\pi\right)^3$, and $\left( |{\bf P}|/m_\pi\right)^3$.
Terms that have been omitted are of the form
$\left(\gamma\rho_d\right)^4 \sim 0.03$, 
$\left( |{\bf P}|\rho_d\right)^4 \sim 0.006$ and 
higher and also 
relativistic corrections of the form $\left(\gamma/M_N\right)^2 \sim 0.002$, or 
$\left( |{\bf P}|/M_N\right)^2 \sim 0.001$ and higher, for nucleon energies of
$E\sim 1\ {\rm MeV}$.
It is interesting to note that contributions from $\Cfourone$ and $\Csixtwo$
occur at N$^3$LO.  However, there are also contributions from the
four-nucleon-one-electric-photon operators in eq.~(\ref{eq:Ldef}) 
with coefficients that exactly reproduce the renormalization scale
independent quantities that occur in eq.~(\ref{eq:LoneRG}).
Therefore, these combinations are higher order than N$^3$LO.
The relative contributions from the P-wave final state interactions
entering at N$^3$LO are much smaller than the $(\gamma \rho_d)^3$
contribution also entering at N$^3$LO.

%%%%%%%%%%  M1  %%%%%%%%%%%%%%%%%%%%%%

The M1-capture  contribution $|\tilde{X}_{M1}|^2$ has been computed
up to NLO (i.e. $Q^1$) for the capture of cold-neutrons
in the theory with pions\cite{SSWst} and $\nopi$\cite{CRSa}.
It is straightforward to extend these results to the capture 
of nucleons with non-zero momentum, and at NLO
we find
\begin{eqnarray}
 |\tilde{X}_{M1}|^2 & = & 
{\kappa_1^2 \gamma^4 \left( {1\over a_1}-\gamma\right)^2 \over
\left({1\over a_1^2} +|{\bf P}|^2\right) \left(\gamma^2 + |{\bf P}|^2 \right)^2
}
\left[
1 + \gamma\rho_d 
- r_0 { \left( {\gamma\over a_1}+|{\bf P}|^2\right)  |{\bf P}|^2 \over
\left({1\over a_1^2} +|{\bf P}|^2\right) \left( {1\over a_1}-\gamma\right)}
- {\Lnp\over\kappa_1}{M_N\over 2\pi}
{ \gamma^2 + |{\bf P}|^2 \over {1\over a_1}-\gamma}
\right]
\ \ \ ,
\label{eq:Mamp}
\end{eqnarray}
where $\kappa_1$ is the isovector nucleon magnetic moment,
$a_1=-23.714\pm 0.013~{\rm fm}$ 
is the scattering length in the $\si$ channel and 
$r_0=2.73\pm 0.03~{\rm fm}$ is the effective range in the $\si$ channel.
The constant $\Lnp$ is a RG invariant combination of parameters,
\begin{eqnarray}
\Lnp & = & 
(\mu-\gamma)(\mu-{1\over a_1})
\left[ \Lone -
{\displaystyle{\pi \kappa_1 \over M_{N}}}
\left( 
{\displaystyle{r_{0}^{(\si)}{}^{{}}
    \over \left( \mu -\frac{1}{a^{(\si)}} \right) ^{2}}}
+
{\displaystyle{\rho _{d} \over (\mu -\gamma )^{2}}}
\right) \right]
\ \ \ ,
\label{eq:Lnpdef}
\end{eqnarray}
that must be determined from data.
It is simplest to determine $\Lnp$  from the cross section for cold neutron
capture, that is dominated by the M1 matrix element.
For incident neutrons  with speed $|v|=2200~{\rm m/s}$ the cross section
for capture by protons at rest is measured to be 
$\sigma^{\rm expt} = 334.2\pm 0.5~{\rm mb}$\cite{CWCa}.
The value of $\Lnp$ required to reproduce a cross section of $334.2~{\rm mb}$
is $\Lnp=-4.513~{\rm fm}^2$ (the experimental uncertainty in this measurement
introduces a negligible uncertainty in our predictions and 
we have used the isopin averaged value of the nucleon mass, $M_N=938.92$).
The largest uncertainty in the M1 contribution to the total cross section 
is expected
to be of the form $|{\bf P}|\gamma\rho_d^2$, which is $\sim 10\%$ at an energy
of $1~{\rm MeV}$
(corrections of the form $\gamma^2\rho_d^2$ have been renormalized away
by fitting $\Lnp$ to the cold neutron capture cross section).  
Given that for energies above $\sim 200~{\rm keV}$ the E1 amplitude dominates
the cross section, a relatively large uncertainty in the M1 amplitude does not 
lead to a large uncertainty in the total cross section.
In fact, we find that by assigning an uncertainty of $3\%$ to the 
E1 cross section and an uncertainty of the form 
$|{\bf P}|\gamma\rho_d^2$ to the
M1 contribution, the uncertainty in the total cross section is 
$\lsim 4\%$ (the maximum uncertainty occuring at an energy of 
$\sim 200~{\rm keV}$), where we have added the errors linearly.
For energies below $8~{\rm keV}$ the uncertainty drops to below $1\%$.
%
%%%%%%%%%%   Figure Cross Sections M1 E1   %%%%%%%%%%%%%%%
%
\begin{figure}[t]
\centerline{{\epsfxsize=4.5in \epsfbox{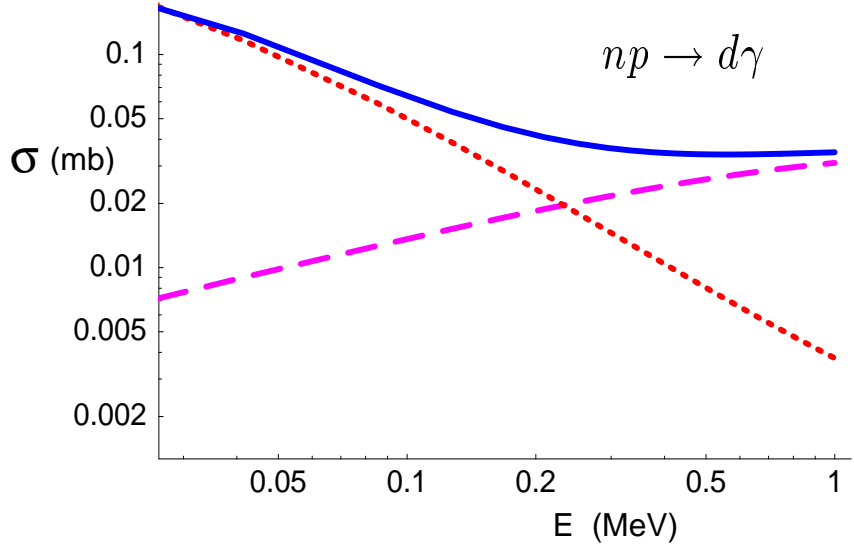}} }
\noindent
\caption{\it
The cross section for $np\rightarrow d\gamma$ as a 
function of the  center-of-mass kinetic energy $E$ in MeV. 
The dotted curve is the contribution from M1-capture, the 
dashed curve is the contribution from E1 capture and the 
solid curve is the sum of the M1 and E1 capture 
cross sections. 
Both the vertical and horizontal axes are logarithmically scaled.
}
\label{fig:M1E1}
\vskip .2in
\end{figure}

At LO, the expression in eq.~(\ref{eq:Mamp}) for the M1 amplitude reproduces
the analogous expression found in ER\cite{ERtheory,Noyes} (after the
typographical errors in \cite{Noyes} have been corrected\cite{CRSa}).  
However, at NLO where the effective range
parameters $\rho_d$ and $r_0$ first appear the expression differs both
qualitatively and quantitatively.
The ER amplitude does not correctly describe physics at distance scales of
order $1/m_\pi$ or shorter.  
In $\nopi$, physics at such distance scales, beyond 
the physics of NN scattering alone, is reproduced by the local counterterm
$\Lnp$.

\begin{table}[!tbp]
\begin{tabular}[h]{ccccc}
\multicolumn{5}{c}{$\sigma (np\rightarrow d\gamma )$}\\ \hline
\multicolumn{1}{c}{  }  
& \multicolumn{3}{c}{ $\nopi$}  
& \multicolumn{1}{c}{ENDF\cite{ENDF}} \\ 
$E$ (MeV) & M1 (mb) & E1 (mb) & M1+E1 (mb) & (mb) \\ \hline
$1.264\times 10^{-8}$ & 334.2 & $5.1\times 10^{-6}$ & $334.2^{\ (*)}$ & 332.0\\
$5.0\times 10^{-4}$ & 1.668 & $1.0\times 10^{-3} $ & 1.669(4) & 1.660\\
$1.0\times 10^{-3}$ & 1.171 & $1.42\times 10^{-3}$ & 1.173(4) & 1.193\\
$5.0\times 10^{-3}$ & 0.496 & $3.17\times 10^{-3}$ & 0.499(4) & 0.496\\
$1.0\times 10^{-2}$ & 0.329 & $4.48\times 10^{-3}$ & 0.333(4) & 0.324\\
$5.0\times 10^{-2}$ & 0.0987 & $9.84\times 10^{-3}$ & 0.109(3) & 0.108\\
0.100 & 0.0501 & 0.0136 & 0.064(2) & 0.0633\\
0.500 & 0.00803 & 0.0260 & 0.034(1) & 0.0345\\
1.00 & 0.00375 & 0.0310 & 0.035(1) & 0.0342
\end{tabular}
\caption{
The cross section for $np\rightarrow d\gamma$ in millibarns
as a function of the 
nucleon center-of-mass energy, $E$.
The counterterm $\Lnp$ is fit to reproduce a cross section of 
$334.2~{\rm mb}$ at an incident neutron speed of
$|{\bf v}|=2200~{\rm m/s}$.
The fact that this cross section is an input is denoted by the 
asterisk$^{\ (*)}$.
The numbers in parenthesis are the uncertainty in the last digit, 
and are estimated
by assigning a fractional error of $(\gamma\rho_d)^4=0.028$ to the 
E1 cross section and 
a fractional error of $\gamma |{\bf P}|\rho_d^2$ to the M1 
cross section.
The last column is the total cross section as extracted from the 
on-line nuclear data center\protect\cite{ENDF}.
}
\label{table1}
\end{table}

A comparison between the cross section obtained with effective field theory
and the numerical values obtained from the 
on-line nuclear data center\cite{ENDF} is shown in table~\ref{table1}.
One sees that the analytic expressions we have obtained reproduces very well 
(within a few percent) the 
numerical values of ref.~\cite{ENDF}.
However, at $E=1~{\rm keV}$ the $\nopi$ cross section is $\sim 2.5\%$
lower than the value from \cite{ENDF}.
The uncertainty in the $\nopi$ value is expected to be 
$\gamma |{\bf P}|\rho_d^2\sim 0.4\%$, and so it is very unlikely
that higher order contributions will bring the 
$\nopi$ value into agreement with the value from ref.~\cite{ENDF} 
at this energy.
A similar, but much weaker statement can be made about the cross section
at $E=10~{\rm keV}$.

%%%%%%%%%%%%  Photo %%%%%%%%%%%%%%%%%%%%%%%%

\begin{figure}[t]
\centerline{{\epsfxsize=4.5in \epsfbox{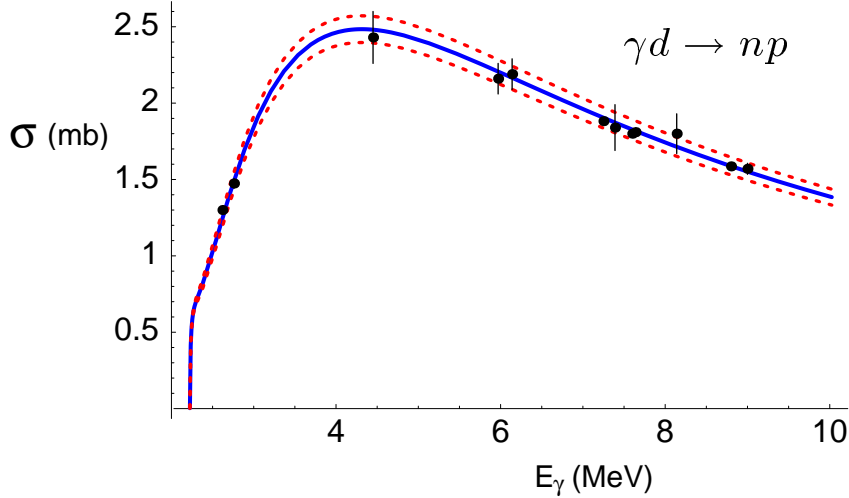}} }
\noindent
\caption{\it
The cross section for $\gamma d\rightarrow np$
as a function of the incident photon energy in MeV. 
The solid curve corresponds to the cross 
section computed in $\nopi$.
The two dotted curves correspond to the uncertainty 
in the $\nopi$ calculation
as estimated in the text.
The data points with error bars can be found 
in table~\ref{table2}.
}
\label{fig:pdhigh}
\vskip .2in
\end{figure}

A measure of the accuracy of our calculation can be determined 
by examining the  deuteron photo-dissociation cross section, which
is related to the capture cross section by
\begin{eqnarray}
\sigma(\gamma d\rightarrow np) & = & 
{2 M_N (E_\gamma - B)\over 3 E_\gamma^2}\ 
\sigma (np\rightarrow d\gamma)
\ \ \ ,
\label{eq:photo}
\end{eqnarray}
where $E_\gamma$ is the incident photon energy and the deuteron is at rest.
A comparison between the low-energy 
cross section computed with $\nopi$ and high precision
experimental values can be seen in table~\ref{table2}.
A detailed and very illuminating 
discussion of the experiments that contributed to these 
data points can be found in ref.~\cite{ASbook}.
In addition, a detailed comparison between the predictions of potential
models, in particular the Bonn r-space potential, with these data
can also be found in \cite{ASbook}.
For the two lowest energy data points
the uncertainties in the $\nopi$ calculation are seen to be nearly a 
factor of two larger than the experimental uncertainties at this order.
%
%%%%%  Figure Photo-Deuteron  threshold %%%%%%%
%
\begin{figure}[t]
\centerline{{\epsfxsize=4.5in \epsfbox{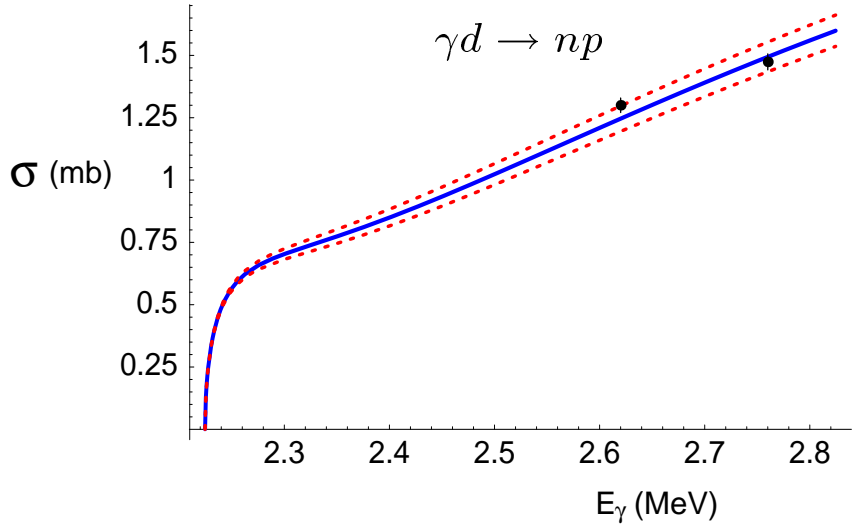}} }
\noindent
\caption{\it
The cross section for $\gamma d\rightarrow np$ 
near threshold,
as a function of the incident photon energy in MeV. 
The solid curve corresponds to the cross section 
computed in $\nopi$.
The two dotted curves correspond to the uncertainty 
in the $\nopi$ calculation
as estimated by the method described in the text.
The two data points with error bars can be found 
in table~\ref{table2}.
}
\label{fig:photodeut}
\vskip .2in
\end{figure}
\begin{table}[!tbp]
\begin{tabular}[h]{ccccc}
\multicolumn{5}{c}{$\sigma (\gamma d\rightarrow np )$}\\ \hline
& \multicolumn{3}{c}{ $\nopi$ }  
& \multicolumn{1}{c}{ expt.\cite{ASbook}}  \\ 
$E_\gamma$ (MeV) & M1 (mb) & E1 (mb) & M1+E1 (mb) &  (mb)\\ \hline
$2.62$ & $0.380$ & $0.866$ & $1.25(5)$ & $1.300\pm 0.029$ \\
$2.76$ & $0.327$ & $1.167$ & $1.50(6)$ & $1.474\pm 0.032$ \\
$4.45$ & $0.128$ & $2.353$ & $2.48(9)$ & $2.43\pm 0.17 $\\
$5.97$ & $0.084$ & $2.122$ & $2.21(8)$ & $2.162\pm 0.099 $\\
$6.14$ & $0.081$ & $2.084$ & $2.17(8)$ & $2.190\pm 0.100 $\\
$7.25$ & $0.066$ & $1.838$ & $1.90(7)$ & $1.882\pm 0.011 $\\
$7.39$ & $0.065$ & $1.808$ & $1.87(7)$ & $1.840\pm 0.150 $\\
$7.60$ & $0.063$ & $1.764$ & $1.83(7)$ & $1.803\pm 0.016 $\\
$7.64$ & $0.062$ & $1.756$ & $1.82(7)$ & $1.810\pm 0.028 $\\
$8.14$ & $0.058$ & $1.656$ & $1.71(6)$ & $1.800\pm 0.130 $\\
$8.80$ & $0.053$ & $1.534$ & $1.59(6)$ & $1.586\pm 0.011 $\\
$9.00$ & $0.052$ & $1.500$ & $1.55(6)$ & $1.570\pm 0.036 $
\end{tabular}
\caption{
The cross section for $\gamma d\rightarrow np$ in millibarns
computed in $\nopi$ and
the experimental values taken from pages 78 and 79 of 
ref.~\protect\cite{ASbook}.
The $\nopi$ cross section is comprised of the E1 amplitude computed to
order N$^3$LO and the M1 amplitude computed to NLO.
The theoretical uncertainties are estimated in the same way as those
in table~\ref{table1}.
}
\label{table2}
\end{table}
A plot of the break-up cross section in Fig.~(\ref{fig:photodeut}),
along with the two low-energy data points,
clearly shows the need for more data in this low-energy region.
A few more high precision measurements between $\sim 2.5~{\rm MeV}$ and 
$\sim 4.0~{\rm MeV}$ would provide important constraints on 
the M1 and E1 amplitudes in the energy region relevant to big-bang
nucleosynthesis.
It does appear that the $\sim 3\%$ uncertainty in the E1 cross section
that we have estimated to arise from
unknown higher order contributions may be an over-estimate.
Fig.~(\ref{fig:pdhigh}) hints that a $1\sigma$ error of $1\%$ or $2\%$
might be appropriate, but this cannot be justified from $\nopi$ alone.
Given that the $\nopi$ at N$^3$LO  
reproduces the E1 amplitude very well, and that there
are only two precise data points in the energy region that is 
sensitive to the M1 amplitude,
it is unlikely that pushing the 
$\nopi$ computation to one higher order would lead to a noticeable
difference in cross section.
Even at this order in the $\nopi$ expansion, we find good agreement
with potential model calculations\cite{ASbook}.
%
%%%   Figure Photo-Deuteron Higher  %%%%%%%
%

Returning to the $np\rightarrow d\gamma$ capture process,
for energies above $\sim 300~{\rm keV}$ the cross section is dominated by 
E1 capture and hence the uncertainty in our calculation
is essentially the uncertainty in the E1 cross section, $\sim 3\%$.
In order to further reduce this uncertainty down to $\sim 1\%$
a N$^4$LO calculation of the E1 amplitude is  required.
As we have discussed previously, at N$^4$LO there is a contribution from a 
local four-nucleon-one-electric-photon interaction that is not 
constrained by 
nucleon-nucleon scattering, but could be determined by the deuteron
photo-disintegration cross section.  
The appearance of such an operator is not
restricted to effective field theory as such interactions will also arise
in potential model calculations of this amplitude, making roughly the same
size contribution. 
An example of this 
is the deuteron quadrupole moment~\cite{mccalc,DanTom}.
To describe the lower energy regime to higher precision, 
the M1 amplitude will
need to be computed to N$^2$LO or higher.
At this order there will be additional counterterms, beyond $\Lnp$ that will
need to be determined by data.
Fig.~(\ref{fig:photodeut}) suggests that the existing two data 
points in this
region  may not be sufficient to achieve this.

%%%%%%%%%  Conclusion  %%%%%%%%%%%%

In conclusion, we have examined the radiative capture process
$np\rightarrow d\gamma$ in the pionless nucleon-nucleon 
effective field theory.
An analytic expression for the cross section in the energy region 
relevant to big-bang nucleosynthesis is presented
and is expected to reproduce the true cross section
at the few percent level.
It is simple to relate the rate for $np\rightarrow d\gamma$ 
to the cross section for $\gamma d\rightarrow np$
and our calculation agrees very well with the existing data.
There is motivation to perform high precision 
measurements of the 
deuteron photo-disintegration cross section at very low energies, to tightly
constrain the M1 contribution to the cross section.

\vskip 0.5in

We would like to thank Gautam Rupak and Scott Burles for 
useful discussions.
We would also like to thank Baha Balantekin, Wick Haxton and 
Brad Keister for bringing this issue to our attention.
This work is supported in part by the U.S. Dept. of Energy under Grants No.
DE-FG03-97ER41014.

\end{document}